# Temperature Dependent Characteristics of Quasi-vertical AlN Schottky Diodes on Bulk AlN Substrate

Md Abdul Hamid[1], Nabasindhu Das[1], Advait Gilankar[1], Brad Lenzen[1], David J. Smith[2], Nidhin Kurian Kalarickal[1]

[1]School of Electrical, Computer, and Energy Engineering, Arizona State University, Tempe, AZ 85281

[2]Department of Physics, Arizona State University, Tempe, AZ 85281

**Correspondence:** mhamid11@asu.edu

**ABSTRACT**

We report on the fabrication and temperature-dependent characterization of MOCVD-grown quasi-vertical AlN Schottky barrier diodes (SBDs) on bulk AlN substrates. The SBDs exhibited high current densities exceeding 2 kA/cm$^2$ at 10 V, with a turn-on voltage of ~3.0 V (at 1 A/cm$^2$) and an on/off ratio >10$^9$ at room temperature. Stable rectifying operation was maintained up to 300 °C (the highest measured temperature), with a pronounced increase in current density at elevated temperatures due to thermally activated carrier transport, accompanied by an increase in extracted Schottky barrier height and a reduction in ideality factor. Capacitance voltage measurements showed strong temperature dependence due to the deep donor nature of Si in AlN, resulting in an increase in the net donor concentration ($N_D-N_A$) from ~5×10$^{17}$ cm$^{-3}$ at 300 K to ~1×10$^{18}$ cm$^{-3}$ at 373 K. Temperature-dependent reverse-bias characteristics were consistent with Poole-Frenkel emission as the dominant leakage mechanism, with an estimated trap energy of ~0.34 eV. Characterization using transmission electron microscopy and energy-dispersive X-ray spectroscopy revealed a ~5 nm AlN$_x$O$_y$ interfacial layer at the metal/semiconductor junction, which likely influences both forward and reverse transport. These results provide insight into carrier transport, leakage mechanisms, interface chemistry, and high-temperature characteristics, and guidance for the future development of high-performance AlN power devices.

**Topics**: Group III-V nitrides, Nanofabrication, Quasi-vertical design, Donor activation, Current Transport

Aluminum nitride (AlN) is an ultra-wide bandgap semiconductor ($E_G \sim 6.2$ eV) with excellent transport properties ($\mu_n \sim 800$ cm$^2$/V-s,)[1], high thermal conductivity (~340 W/m-K)[2], and excellent thermal stability (melting point ~2100 °C). [3,4] Its large critical electric field (~15 MV/cm) makes AlN a promising candidate for next-generation power electronic devices targeting applications in electric vehicles, data centers, motor drives, and smart power grids. [5–7] In addition, the direct bandgap of AlN enables deep-ultraviolet (DUV) light emission, extending its utility to photonic devices operating beyond the conventional UV-visible spectrum. [8–10] Owing to its intrinsic radiation hardness and high-temperature stability, AlN is also well suited for operation in harsh environments, including high temperature and high radiation systems. [11,12]

Recent advances in the development of large area substrates, have enabled realization of high-quality epitaxial growth directly on bulk AlN platform.[13–15] Low defect density AlN epilayers can now be achieved using metal organic chemical vapor deposition (MOCVD) under high-temperature growth conditions.[3-5] Furthermore, controlled n-type conductivity with mobility exceeding 400 cm$^2$/V-s and compensating acceptor concentration below $1\times10^{17}$ cm$^{-3}$ has been demonstrated in Si-doped AlN.[16] Recent efforts have also demonstrated functional AlN devices including Schottky barrier diodes (SBDs) and field effects transistors (FETs) with breakdown voltage in the kV range and current density in the multi-kA/cm$^2$ range (only SBDs). *C. E. Quiñones* *et al.* have reported AlN SBDs with an ideality factor <1.2, a current density > 5kA/cm$^2$, and a low breakdown voltage of 680 volts.[17] An improved breakdown voltage of 1.15 kV was reported for AlN SBDs by the formation of an oxide layer via rapid thermal annealing,[18] and AlN SBDs were reported with breakdown up to 3.0 kV using an UID GaN Cap layer.[19] The high temperature characteristics of AlN-based Schottky barrier diodes incorporating a GaN cap layer at the metal/AlN interface, which is likely to influence the overall device performance, were also reported.[20] These reports demonstrate promising AlN-based device performance; however, further research in areas such as carrier transport, dopant compensation, and temperature-dependent characteristics is still required in order to realize the full theoretical performance of AlN power devices. This article presents a detailed study of the Ni/AlN quasi-vertical Schottky barrier diodes fabricated on bulk AlN substrates, with a particular focus on its temperature-dependent behavior including I-V, capacitance and breakdown characteristics, helping realize the full potential of AlN devices.

The epitaxial AlN and AlGaN layers depicted in **FIG. 1(a)** were grown on an AlN bulk substrate using the metal organic chemical vapor deposition (MOCVD) method. An n-doped (Si:2×10$^{19}$ cm$^{-3}$) Al$_{0.8}$Ga$_{0.2}$N buffer layer was grown first on the insulating AlN substrate to serve as the bottom contact layer. A 100 nm thick Si-doped (Si:2×10$^{19}$ cm$^{-3}$) graded Al$_x$Ga$_{1-x}$N layer was grown on top to mitigate the band offset between Al$_{0.8}$Ga$_{0.2}$N and AlN, as shown in **FIG. 1(b)**. A 1-micron thick AlN drift layer with a Si doping concentration of 1×10$^{18}$ cm$^{-3}$ was grown above the graded Al$_x$Ga$_{1-x}$N layer. The AlN epilayer was capped with a 100 nm thick reverse-graded (Si:2×10$^{19}$ cm$^{-3}$) Al$_x$Ga$_{1-x}$N (x=1 to 0.3) layer. High-resolution X-ray diffraction (XRD) was conducted to confirm the epitaxial stack and Al compositions, as shown in **FIG. 1(c)**.

For device fabrication, the top 100 nm of the reverse-graded $Al_{0.3}Ga_{0.7}N$ layer and 20 nm of the AlN epilayer were etched away using a chlorine-based ($BCl_3/Cl_2/Ar$) inductively coupled plasma reactive ion etching (ICP-RIE) process. Then, 1.3-micron deep mesas were defined and etched out using a $SiO_2$/Ni bilayer hard mask. The etch process included a fast etch rate ($BCl_3/Cl_2/Ar$) ICP-RIE process to remove the AlN layer and the graded AlGaN layer, followed by a slower ICP-RIE process to remove an additional 200 nm from the n++ $Al_{0.8}Ga_{0.2}N$ contact layer. This two-step process was used to minimize etch damage to the $Al_{0.8}Ga_{0.2}N$ contact layer. The exposed surface of the $Al_{0.8}Ga_{0.2}N$ contact layer was cleaned using oxygen ashing, followed by a 1-minute dip in hydrochloric (HCl) acid. To form the ohmic contacts, a metal stack of Ti/Al/Ni/Au (20/120/30/50 nm) was deposited via e-beam evaporation. In addition to the Ti-based ohmic stack, we also deposited Cr/Al/Ni/Au (20/120/30/50 nm) and V/Al/V/Au (15/80/20/100 nm) metal stacks for comparison. The ohmic metallization was carried out by annealing the stack in $N_2$ at 950°C for 30 seconds. Finally, the metal stack Ni/Au (60/30 nm) was deposited on the exposed AlN epilayer to form the Schottky contact. The fabricated diodes had an anode diameter (D) of 50 microns. The spacing between the Schottky metal and the edge of the mesa was kept constant (2 µm) for all the devices. The scanning electron microscope (SEM) image in **FIG. 1(d)** shows the top view of an SBD with D=50 µm and W=5 µm. A cross-sectional high-resolution TEM image in **FIG. 1(e)** of the Ni-Schottky interface shows the presence of a 5 nm Aluminum oxynitride ($AlN_xO_y$) layer as identified by energy-dispersive X-ray spectroscopy. This thin interlayer likely formed due to oxidation of the exposed AlN surface. The EDAX profiles in **FIG. 1(f)** show atomic percentage of 68.7% for aluminum, 16.6% for oxygen, and 14.7% for nitrogen in the interlayer.

**FIG. 1**

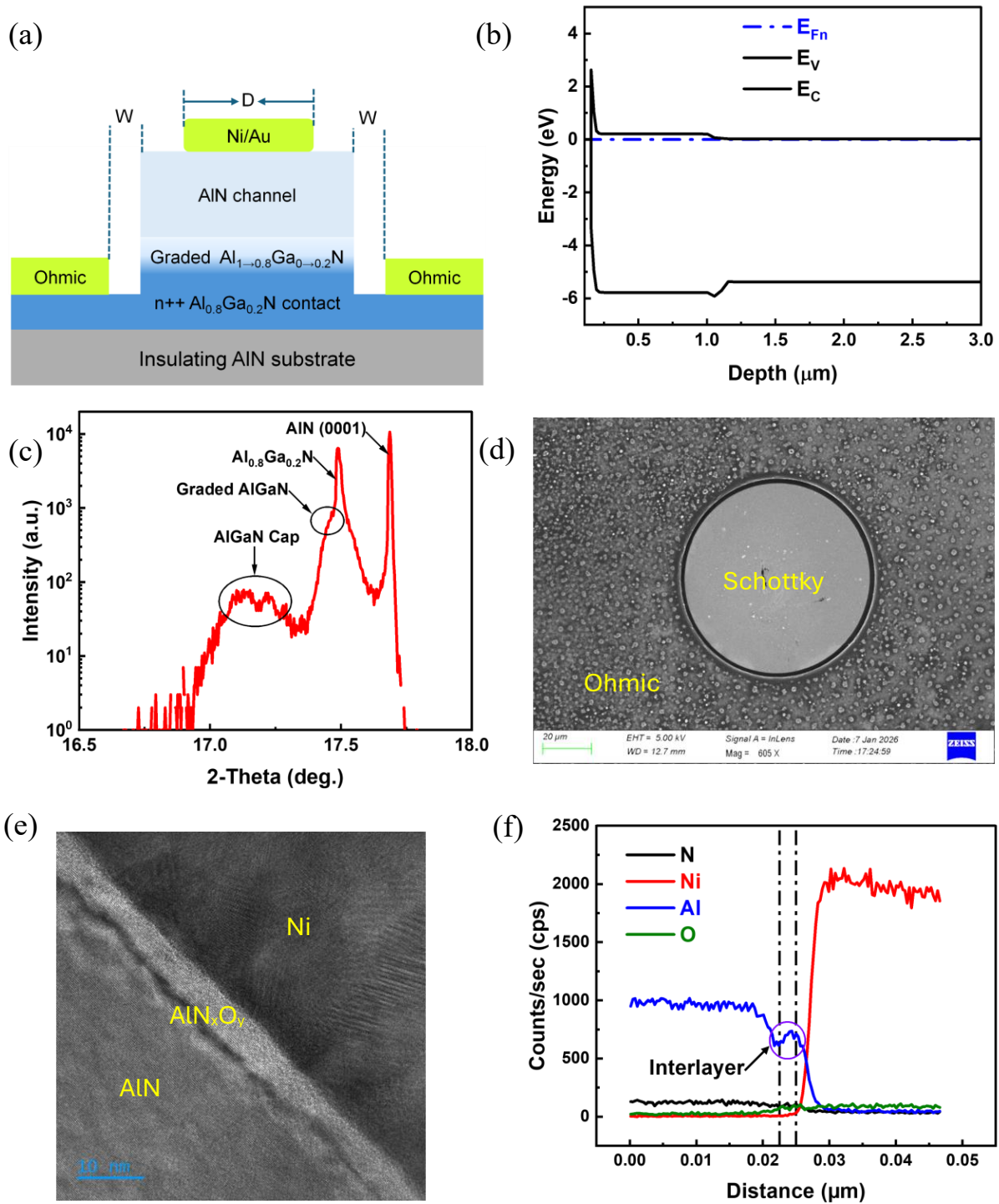

(a) Device schematic of AlN quasi-vertical Schottky barrier diodes, (b) Simulated band diagram assuming a Schottky barrier height of 2.5 eV, (c) High-resolution x-ray diffraction profile of the AlN epilayer. The graded AlGaN cap layer is etched away before fabricating SBDs (d) SEM top view of AlN SBDs, (e) High resolution TEM cross section of AlN/Ni interface, and (f) EDAX profile showing composition of Al, O and N across interface.

**FIG. 2(a)** shows the forward current-voltage (I-V) characteristics in both linear and log scale for an AlN SBD with an anode diameter of 50 μm. Neglecting the series resistance, the forward thermionic I-V relation can be expressed as,

$$I = I_S \left( e^{\frac{qV}{nkT}} - 1 \right) \tag{1}$$

In equation (1), n is the ideality factor, and $I_S$ is the reverse saturation current, which can be expressed as,

$$I_S = AA^* T^2 e^{\frac{-q\Phi_B}{kT}} \tag{2}$$

Where A is the anode area, A* is Richardson's constant, and $\Phi_B$ is the effective Schottky barrier height. For AlN devices, A* is calculated to be 57.68 A/cm$^2$.K$^2$, considering m*=0.48m$_o$.[21] The fabricated SBDs exhibited a turn-on voltage of around ~3.0 volts (measured at 1 A/cm$^2$) and a very low leakage current in the range of ~1.0×10$^{-6}$ A/cm$^2$ (limited by the noise floor of the measurement setup). The diodes displayed high forward current density in the kA/cm$^2$ range and a high ON/OFF ratio of ~10$^9$, exceeding some previous reports at room temperature.[21-24] SBDs with anode diameter of 50 μm show current density exceeding 2 kA/cm$^2$ at 10 V. The ON resistance in **FIG. 2(b)** is determined to be ~10 mohm.cm$^2$ at 4 V and below 1 mohm.cm$^2$ at 10 V. The ideality factor (η) deviates from unity and is extracted to be ≥ 2.2 (the lowest value) at room temperature. This indicates a significant deviation from the ideal thermionic emission behavior, likely due to non-ideal current-transport mechanisms, such as barrier inhomogeneity and interface-trap emission. As shown in **FIG. 1(e)**, a thin AlN$_x$O$_y$ interlayer is present between the Ni Schottky contact and the AlN epilayer. This interlayer likely presents a tunnelling barrier for electrons under forward bias.

Due to the defective nature, transport through the $AlN_xO_y$ interlayer will be mediated by defects which results in the high η values. Due to the high η, we estimate an apparent barrier height of ~1.37 eV at 300 K, which is comparable to some previous reports.[22,25] To determine the Schottky barrier height accurately, it is necessary to obtain ideality factors closer to unity.[26] In addition to the high ideality factor, we also observed the presence of kinks in the sub-threshold region of the I-Vs for a majority of the measured devices. The exact origin of these kinks remains unclear; however, charge transport through the $AlN_xO_y$ interlayer is a likely contributing factor.

**FIG. 2**

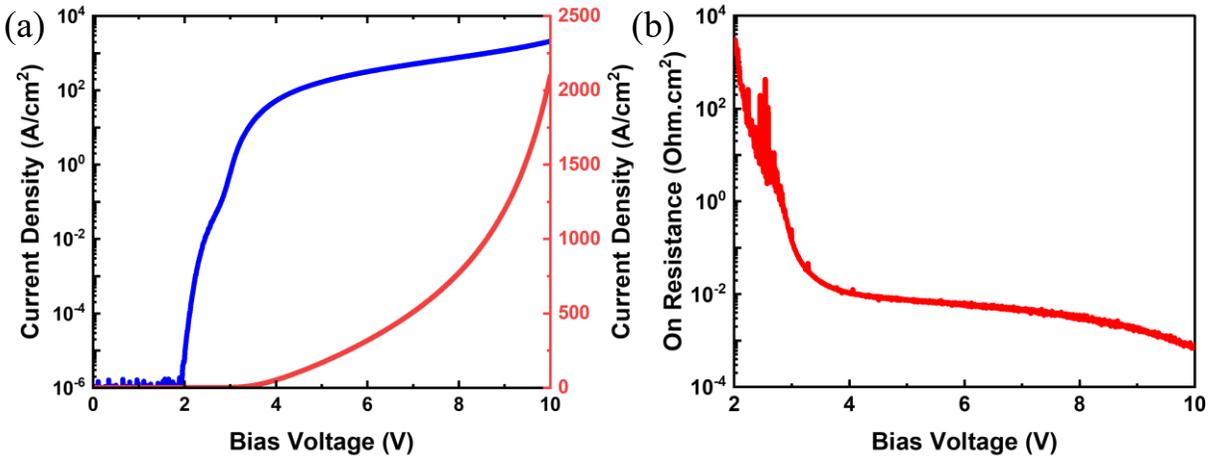

(a) Linear and log-scale I-V characteristics of 50 µm diameter AlN Schottky barrier diodes

(b) ON resistance ($R_{on}$) vs forward bias voltage.

To characterize the ohmic contacts, circular transfer length measurement (CTLM) structures were fabricated on the n++ 80% AlGaN contact layer with an inner electrode diameter of 100 microns. **FIG. 3(a)** shows the IV characteristics of the CTLM structures (Ti/Al/Ni/Au) with spacing ($L_{SP}$) of 6 to 20 µm. The contacts display Schottky behavior with a turn-on of around 2 V, even after annealing at 950°C in $N_2$, suggesting a large barrier for electron flow. The contacts also did not display a linear increase in resistance (after turn-on) with spacing (**FIG. 3(a)**). This is likely due to the total resistance between the two CTLM pads being limited largely by the contact resistance and not by the low sheet resistance of the n++ $Al_{0.8}Ga_{0.2}N$ contact layer. The absence of a monotonic trend in resistance suggests significant inhomogeneity in the ohmic contacts. Reverse graded AlGaN contacts are likely to be needed to solve this issue. **FIG. 3(b)** shows a comparison between the three contact stacks (Ti-based, Cr-based, and V-based) for a CTLM spacing of 10 microns. No significant difference is observed among the three contact stacks in terms of I-V or differential resistance, suggesting that the barrier to electron injection is likely the high affinity of the high-composition $Al_{0.8}Ga_{0.2}N$ layer. **FIG. 3(c)** shows the TLM comparison between the metal stacks at 473K. The contacts display no significant improvement at 473K, displaying similar Schottky-like turn-on behavior. However, the V-based contact stack shows higher resistance at 473K, compared to the Ti- and Cr-based stacks.

**FIG. 3**

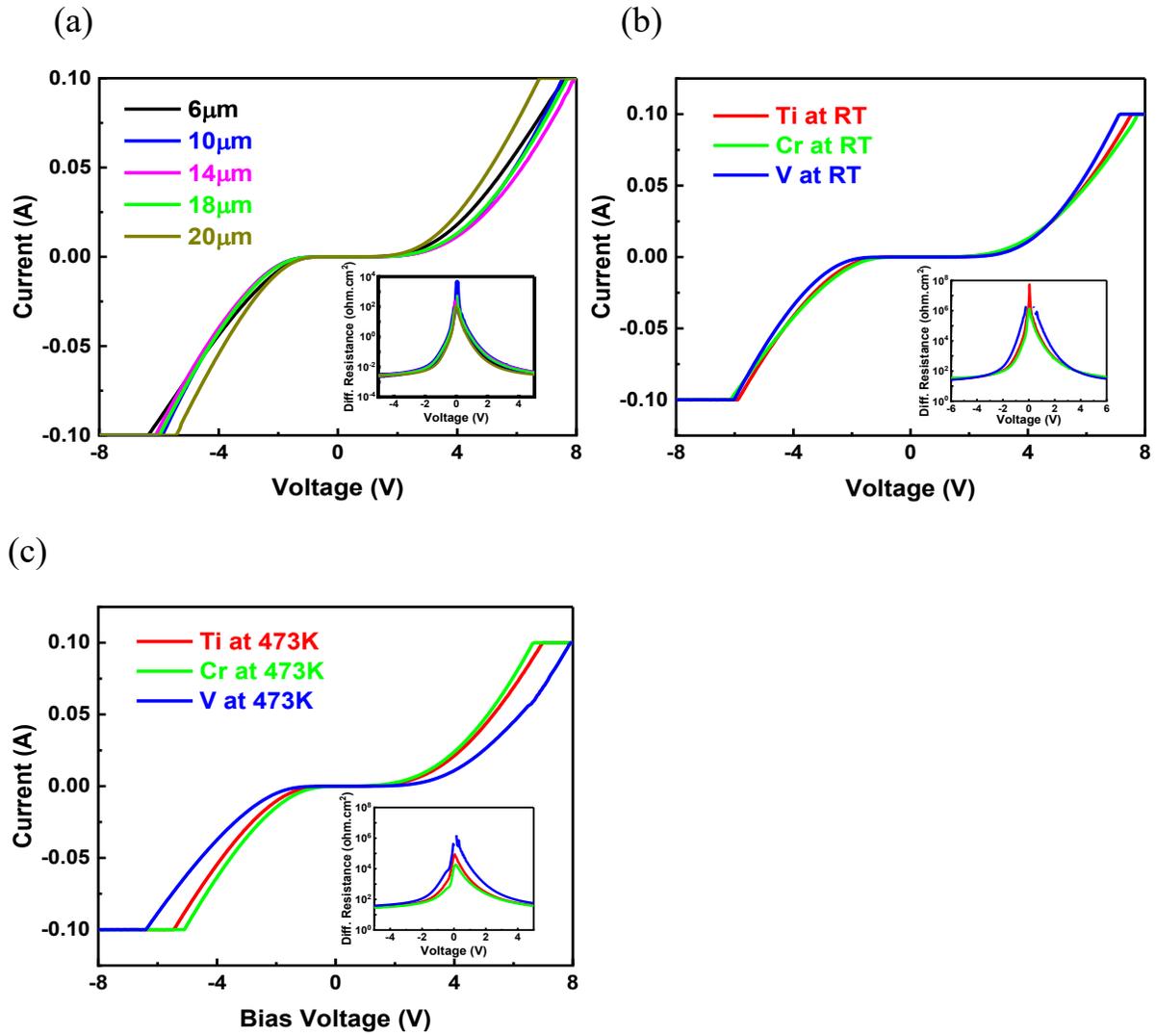

I-V measurements taken on circular TLM structures with (a) Ti/Al/Ni/Au contact at 300 K for different values of TLM spacing, (b) comparison of CTLM characteristics measured for Ti-based, Cr-based, and V-based contacts at a spacing of 10μm at 300 K, (c) at 473K. The insets in each figure show the differential resistance normalized to the inner TLM area (100-micron circle).

**FIG. 4(a)** shows the capacitance-voltage profile measured at 300K using a 150 μm Schottky pad. No significant frequency dispersion was observed between 10 kHz and 1 MHz. **FIG. 4(a)** and **FIG. 4(b)** shows the $1/C^2$ vs voltage and the apparent donor concentration ($N_D$-$N_A$) vs depth profiles, respectively. At 300 K, $N_D$-$N_A$ value of ~$5\times10^{17}$ cm$^{-3}$ is measured suggesting potential compensation of Si donors. Furthermore, an unreasonably high built-in voltage ($V_{bi}$) of ~9.4 V is extrapolated from the intercept of the $1/C^2$ vs voltage plot.

**FIG. 4**

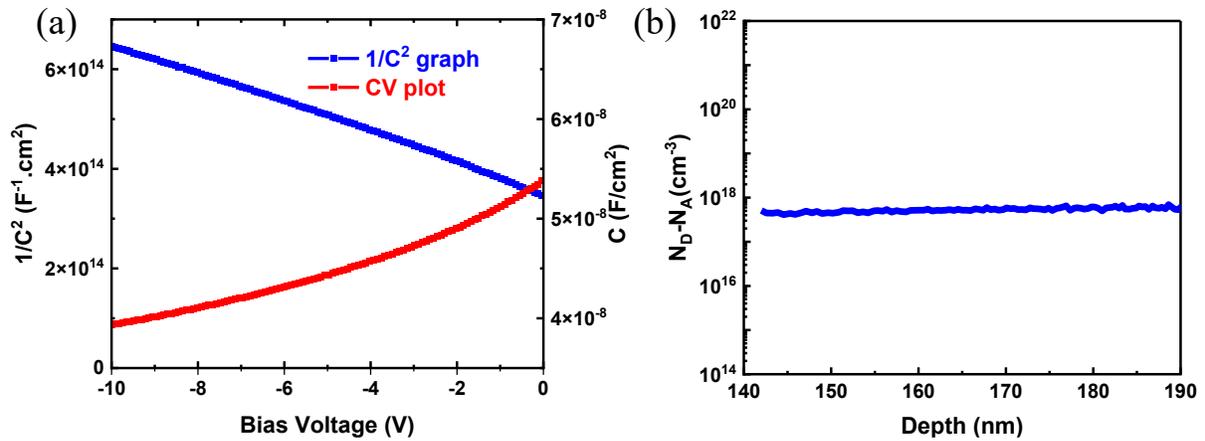

(a) Capacitance (C) and $1/C^2$ vs bias voltage relationship at 100kHz (150 μm diameter Schottky pad) at 300 K, (b) Apparent net donor concentration vs depth extracted from C-V at 300 K.

To determine the stability and robustness of these AlN SBDs, current-voltage and capacitance-voltage characteristics were measured from room temperature up to 573 K. The I-V characteristics shown in **FIG. 5(a)** exhibited stable operation of the AlN SBDs up to 573K. An on/off ratio exceeding $10^7$ is measured even at 573 K. As expected, the SBDs exhibit significantly higher forward current density at higher temperatures due to enhanced carrier activation. The effective turn-on voltage (measured at 1.0 A/cm$^2$) of the SBDs also displays a significant reduction from 3.0 V at 300 K to 2.29 V at 573 K. At high temperatures (>500 K), no kinks were observed in the sub-threshold region of the forward I-V. **FIG. 5(b)** shows the temperature dependence of the ideality factor (η) and apparent Schottky barrier height ($\varphi_{eff}$) up to 573K. η decreases from 3.6 at 300 K to 2.07 at 573 K. In addition, the apparent Schottky barrier height ($\varphi_{eff}$) increases from 1.2 eV at room temperature to 1.93 eV at 573K. As explained before, the higher ideality factor at room temperature is likely due to the presence of the thin $AlN_xO_y$ interlayer between the Ni contact and the AlN layer. At higher temperatures, electrons can overcome this tunnelling barrier because of the higher thermal energy, resulting in a lower ideality factor.

**FIG. 5**

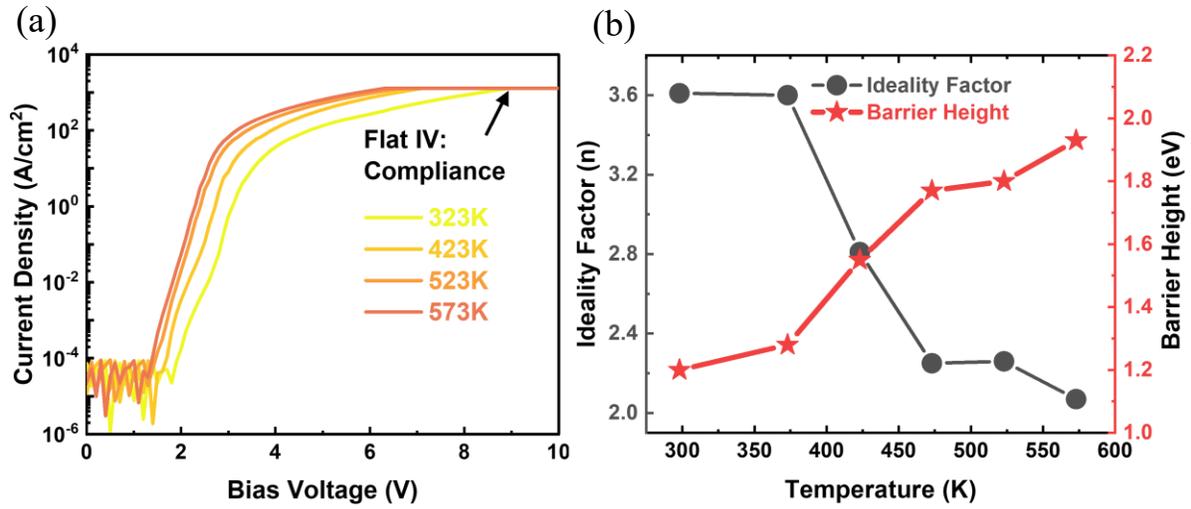

(a) Temperature-dependent forward I-V characteristics of AlN SBDs, (b) Temperature dependence of ideality factor (η) and apparent Schottky barrier height ($\phi_{eff}$).

**FIG. 6**

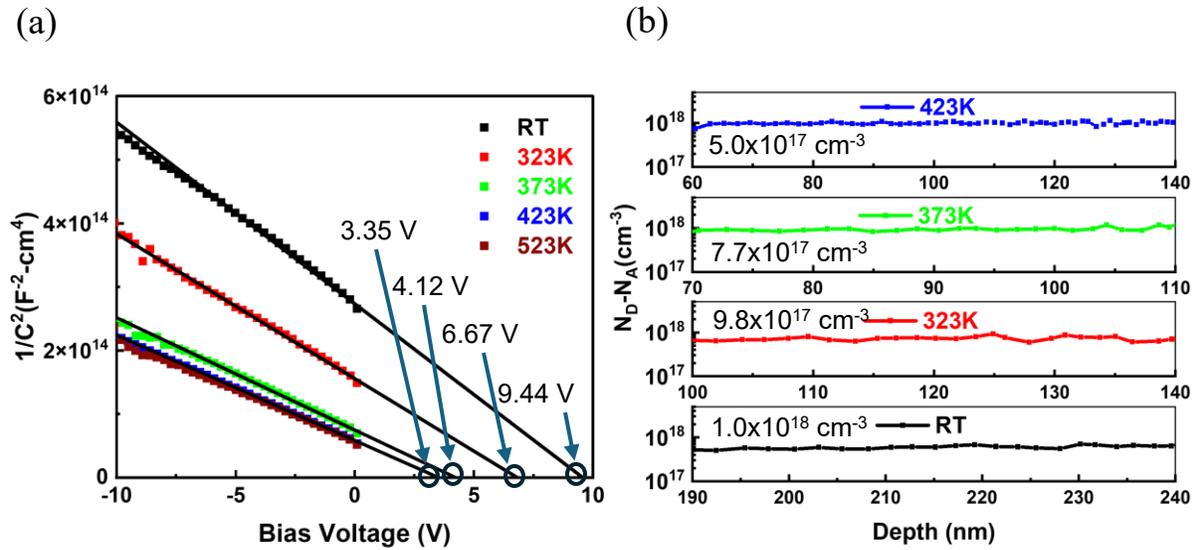

(a) Temperature dependence of $1/C^2$ vs Voltage profile with the extracted $V_{bi}$, (b) Temperature dependence of net donor concentration ($N_D$-$N_A$).

**FIG. 6(a)** shows the temperature dependence of the net donor concentration ($N_D$-$N_A$) and the built-in voltage extracted from capacitance-voltage measurements. The $N_D$-$N_A$ value is found to increase from ~$5\times10^{17}$ cm$^{-3}$ at 300 K to ~$1\times10^{18}$ cm$^{-3}$ at 373 K. No further increase in $N_D$-$N_A$ value is observed beyond 373K with temperature. This temperature dependence is likely due to the deep-level nature of Si donors in AlN. During C-V measurements, the edge of the depletion region in AlN is modulated by the small signal AC bias. The donors (deep level) in this modulated region respond by emitting electrons in the negative half cycle and capturing electrons in the positive half cycle of the AC signal. If the emission rate is slow not all donors respond to the AC signal, resulting in a lower apparent capacitance (and $N_D$-$N_A$ value). All Si donor atoms respond due to the higher emission rates at higher temperatures, resulting in a higher capacitance (and $N_D$-$N_A$ value). The $N_D$-$N_A$ value at high temperature approaches the expected Si concentration in the AlN epilayer, suggesting that the compensating acceptor concentration is at least an order of magnitude lower. The built-in voltage ($V_{bi}$) extrapolated from the $1/C^2$-Voltage plot also shows significant temperature dependence, dropping from 9.44 V at 300 K to 3.35 V at 523 K.

**FIG. 7(a)** shows the reverse-bias characteristics of AlN SBDs measured using flourinert FC-40. The SBDs exhibited repeatable and stable reverse I-V characteristics up to -200 V. Assuming a breakdown current limit of 0.1 A/cm$^2$, a breakdown voltage of ~200 V is estimated, which corresponds to a parallel plane junction field of 6.4 MV/cm (assuming $N_D$-$N_A$=$5\times10^{17}$ cm$^{-3}$). To understand field-crowding effects, TCAD 2-D simulations were performed on the Silvaco Victory Device platform. As shown in **FIG. 7(b)**, a peak electric field of ~13 MV/cm is observed at the edge of the anode contact due to the lack of field termination. Effective field-management structures, such as field plates and mesa-etch termination, could be used to further enhance the junction breakdown field. **FIG. 7(c)** shows the temperature-dependent reverse leakage up to -150

V. The leakage increases by one order of magnitude from $10^{-2}$ A/cm$^2$ to $10^{-1}$ A/cm$^2$ (at -150 V) as the temperature increases from 288 K to 448 K.

The reverse I-V characteristics have been further analyzed by fitting with the Poole-Frenkel (field-enhanced trap emission) transport model. As shown in **FIG. 7(a)**, an excellent linear fit is obtained for the ln(J/E) versus $E^{1/2}$ graph, suggesting the leakage current is dominated by field-enhanced trap emission. **FIG. 7(d)** shows the plot of ln(J/E) versus 1000/T at different electric fields, yielding an average effective trap level of 0.34 eV below the conduction band. The extracted trap energy levels are similar across all electric fields, indicating that Poole-Frenkel is likely the dominant transport mechanism. The electric fields used in this analysis are parallel-plane 1-D fields. However, as shown in **FIG. 7(b)**, the electric field at the anode edge is significantly higher, and the edges likely serve as the primary pathway for reverse leakage current. Consequently, the trap energy extracted using the 1-D approximation is an underestimation of the true trap energy level.

**FIG. 7**

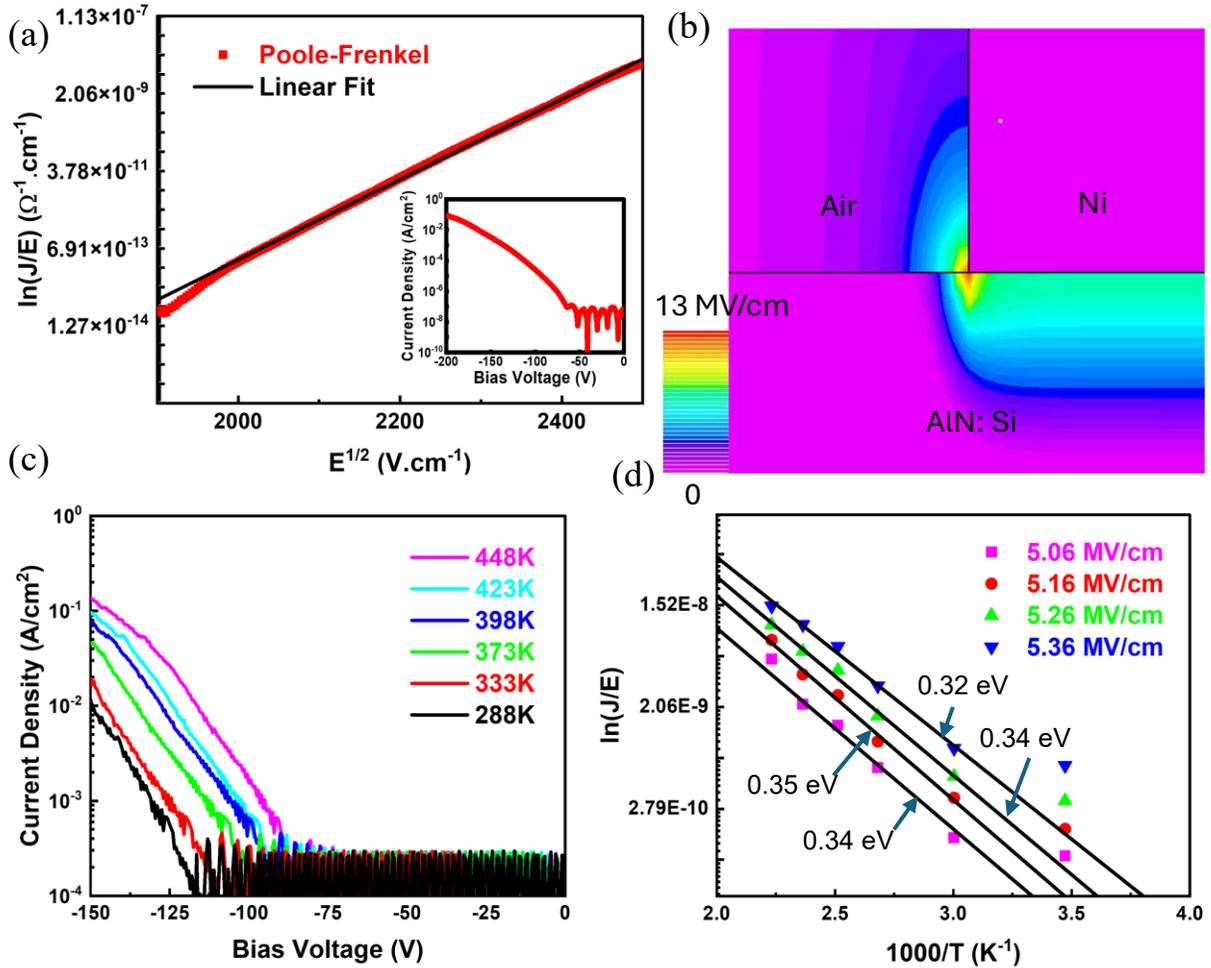

(a) Poole-Frenkel fit (ln(J/E) vs $E^{1/2}$) and reverse bias (inset fig.) of AlN SBDs, (b) Simulated field profile at a reverse bias of -200 V showing peak electric field of 13 MV/cm at the anode edge, (c) Temperature dependence of reverse leakage current (d) ln(J/E) vs 1000/T graph at different electric fields with extracted $\varphi_{eff}$.

In conclusion, we have demonstrated quasi-vertical AlN Schottky barrier diodes on bulk AlN substrates and studied the temperature-dependent characteristics. The diodes show high forward current density exceeding 2 kA/cm$^2$ and a turn-on voltage of 3.0 volts. The SBDs display a high ideality factor of 3.6 at room temperature, which decreases to 2.07 at 573 K, indicating the presence of non-ideal transport/recombination mechanisms. The depletion capacitance, net donor concentration and built in voltage all display temperature dependence, suggesting partial response from deep Si donor levels near room temperature. The SBDs also display non-destructive and repeatable reverse-bias measurements up to -200 volts at room temperature. The reverse-bias leakage increases at higher temperatures, consistent with the Poole-Frenkel field-enhanced trap emission model. Temperature dependence of reverse bias leakage measurements has provided an average effective trap level of 0.34 eV below the conduction band. This comprehensive study of AlN Schottky barrier diodes provides an in-depth overview of device performance at room and elevated temperatures.

## ACKNOWLEDGMENTS

The work in ASU NanoFab was supported in part by the National Science Foundation award ECCS-2025490. We acknowledge the use of facilities within the Eyring Materials Center at Arizona State University supported in part by NNCI-ECCS-1542160. We also thank Professor Martha McCartney and Dr. Manuel Roldan-Gutierrez for their assistance with microscopy and spectroscopy.

## AUTHOR DECLARATION

**Conflict of Interest**

The authors have no conflicts of interest to disclose

**Author Contributions**

**Md Abdul Hamid:** Conceptualization (supporting), Data acquisition & curation (lead), Formal analysis, Visualization, Writing – original draft & editing; **Nabasindhu Das:** Data acquisition (supporting); **Advait Gilankar:** Data acquisition (supporting); **Brad Lenzen:** Data acquisition (supporting); **David Smith:** Data acquisition and visualization (supporting); **Nidhin Kurian Kalarickal:** Supervision, Conceptualization (lead), Methodology, Funding acquisition, Writing – review

## DATA AVAILABILITY

The data that support the findings of this study are available from the corresponding author upon reasonable request.